 \documentstyle[11pt]{article}

\begin{document}

\thispagestyle{empty}

\vglue 1cm

\noindent {\Large \bf Fractional supersymmetry 
and hierarchy \\ of shape invariant potentials}

\vspace{1cm}

\noindent {\bf M. Daoud} \footnote{Permanent address: 
D\'epartement de Physique, Facult\'e des Sciences,
Universit\'e Ibn Zohr, Agadir, Morocco} 
\footnote{Electronic mail: m\_daoud@hotmail.com} 
and 
{\bf M.~R. Kibler} 
\footnote{Electronic mail: m.kibler@ipnl.in2p3.fr}

\vspace{1.5cm} 
\noindent Universit\'e de Lyon, \\
IPNL, universit\'e Lyon~1, CNRS/IN2P3, \\
Villeurbanne, F-69622, France 


 \vspace{2cm}

\noindent Fractional supersymmetric quantum mechanics is developed from a 
generalized Weyl-Heisenberg algebra. The Hamiltonian and the supercharges of 
fractional supersymmetric dynamical systems are built in terms of the 
generators of this algebra. The Hamiltonian gives rise to a hierarchy of 
isospectral Hamiltonians. Special cases of the algebra lead to dynamical 
systems for which the isospectral supersymmetric partner Hamiltonians are 
connected by a (translational or cyclic) shape invariance condition.

\newpage


\section{INTRODUCTION}


Supersymmetry was initially introduced in high energy physics, as a kind 
of symmetry between bosons and fermions, to describe fundamental interactions 
of Nature in an unified way (e.g., see Ref.~1). 
Supersymmetry cannot be an exact symmetry. In order to understand supersymmetry 
breaking in quantum field theory, Witten studied 
supersymmetric quantum mechanics (SUSYQM).$^2$ In the present days, SUSYQM 
turns out to be a powerful tool to investigate integrability 
in quantum mechanics.$^{3-5}$ In this connection, the concept of shape invariant
potential was introduced by Gendenshtein.$^6$ This concept is especially useful 
to determine the spectrum of exactly solvable potentials. Indeed, for a given 
solvable potential, shape invariance implies integrability. It is now 
well-known there are three kinds of shape invariant potentials, namely, 
translational shape invariant potentials,$^{7,8}$ scaling shape invariant 
potentials$^{9,10}$ and cyclic shape invariant potentials.$^{11,12}$

For any exactly solvable Hamiltonian (shape invariant or not), SUSYQM 
provides us with a process to generate a supersymmetric 
partner Hamiltonian. This process can be used successively to span a hierarchy 
of isospectral Hamiltonians.$^5$

The aim of this work is to study shape invariant 
potentials together with the generation of a 
hierarchy of isospectral superpartner Hamiltonians in the 
framework of {\it fractional} SUSYQM of order 
$k$ ($k = 3, 4, \cdots$).

In general, to pass from {\it ordinary} SUSYQM to {\it fractional} SUSYQM of 
order $k$ (abbreviated as $k$-SUSYQM in the following), it is necessary to 
replace the $Z_2$-grading of the relevant Hilbert space by a $Z_k$-grading. This 
amounts either to replace a fermionic degree of freedom by a 
para-fermionic$^{13-18}$ one, of order $k-1$, or to introduce 
$k$-fermions,$^{19-22}$ which are objects interpolating between bosons and
fermions. Quantum groups, with the deformation parameter being a root of unity, 
play also an important role in the development of $k$-SUSYQM.$^{23,31}$ On 
the other hand, a realization of bosonized $k$-SUSYQM can be 
developed owing to the introduction of a Klein operator of order 
$k$ ($K^k = 1$) which induces a $Z_k$-grading.$^{18,32}$ In this direction, 
a relation exists between $k$-SUSYQM and hidden supersymmetric 
structures.$^{33-36}$

The approach of $k$-SUSYQM developed in the present paper took its origin 
in Ref.~22 (see also Refs.~37-39 for some similar developments). It is based on 
a $Z_k$-graded Weyl-Heisenberg algebra $W_k$.  Section 3 deals 
with algebra $W_k$ and its use for generating a family of $k$ isospectral 
Hamiltonians. In Sections 4 and 5, some specific Hamiltonians (with 
translational shape invariant potentials or cyclic shape invariant potentials) 
corresponding to particular cases of the algebra $W_k$ are studied. We will 
start in Section 2 with some preliminaries and motivations. 

Throughout the present paper, $[A,B]$ and $\{A,B\}$ stand for the commutator 
and the anti-commutator of the operators 
$A$ and $B$, respectively. The operator $A^{\dagger}$ 
denotes the adjoint of $A$. The symbol $\delta$ is the Kronecker delta. 
Many quantities are defined modulo $k$ 
($\Pi_k \equiv \Pi_0$, $H_k \equiv H_0$, $F_k \equiv F_0$ and 
$V_k \equiv V_0$). As usual, $f \circ g (x) = f(g(x))$ for two functions $f$ 
and $g$. We shall use the convention according to which 
$\sum_{i=a}^{b} x(i) = 0$ when $b < a$ and the symbols 
$S_0$ and 
$S_1$ for denoting the sets 
$\{0, 1, \cdots, k-1\}$ and 
$\{1, 2, \cdots, k\}$, respectively.


\section{PRELIMINARIES AND MOTIVATIONS}


For the purpose to establish our notations and to present our motivations, we 
shall begin with a brief review of {\it ordinary} SUSYQM, corresponding to 
$k = 2$, and of shape invariance (for more details, see Refs.~3-5).


\subsection{Ordinary Supersymmetric Quantum Mechanics}


Let us start with ordinary SUSYQM. A supersymmetric
dynamical system is defined by a triplet $(H, Q_+, Q_-)_2$ of linear operators 
acting on a $Z_2$-graded Hilbert space ${\cal H}$ 
and satisfying the following relations
$$
H   =   H^{\dagger}, \quad
Q_- = Q_+^{\dagger}, \quad Q_{\pm}^2 = 0,
$$
$$
\{Q_- , Q_+\} = H, \quad [H , Q_{\pm}] = 0.
$$
The operators $Q_+$ and $Q_-$ are the 
supercharges of the system. The self-adjoint operator 
$H$, the supersymmetric Hamiltonian for the (one-dimensional) system, 
can be written as 
$$
H = H_0 + H_1,
$$
where $H_0$ and $H_1$ act on the states $\vert n,0 \rangle$ and 
                                       $\vert n,1 \rangle$
of grading $0$ and $1$, respectively. These states span the Hilbert space 
$$
{\cal H} = \{\vert n,s \rangle: n \ {\rm ranging;} \ s = 0, 1\}.
$$
We shall assume that there is no 
supersymmetry breaking. In this situation, the 
Hamiltonians $H_0$ and $H_1$ are isospectral except that the ground state of 
$H_0$ has no supersymmetric partner in the spectrum of $H_1$.

Now suppose that $H_0$ has $p$ states $\vert n,0 \rangle$ with 
$n = 0, 1, \cdots, p-1$ ($p \ge 2$). From 
the Hamiltonian $H_1$ 
with $p-1$ states $\vert n,1 \rangle$ ($n = 1, 2, \cdots, p-1$), 
we can find a supersymmetric partner $H_2$ 
with $p-2$ states $\vert n,2 \rangle$ ($n = 2, 3, \cdots, p-1$) 
and we can repeat this 
process to generate a hierarchy of $p$ Hamiltonians 
$H_0$, $H_1$, $\cdots$, $H_{p-1}$. The Hamiltonian $H_m$ ($0 <   m <   p$) has 
the same energy spectrum than $H_0$ except that the $m-1$ first energies 
of $H_0$ do not occur in the spectrum of $H_m$. This result remains valid when 
$p$ goes to infinity.

We can then ask the following question. What happens when we go from ordinary SUSYQM to 
$k$-fractional SUSYQM (with $k = 3, 4, \cdots $)? We shall answer this question 
by showing that a hierarchy of isospectral Hamiltonians 
$H_0$, $H_1$, $\cdots$, $H_{k-1}$ can be constructed from a single Hamiltonian 
$H_0$ by making use of a $Z_k$-graded Weyl-Heisenberg algebra. This construction 
shall be achieved without a repetition process of the type 
$H_0 \to H_1$, $H_1 \to H_2$, $\cdots$, $H_{k-2} \to H_{k-1}$ 
as used in ordinary SUSYQM.


\subsection{Shape Invariance}


It is known that there exists a set of exactly solvable potentials 
characterized by an integrability condition known as shape invariance 
condition.$^{3,5,6}$ In connection with ordinary SUSYQM, this shape 
invariance condition leads in an easy way to the spectrum of any invariant 
shape potential. 

More precisely, let us consider the partner potentials $V_0(x,a_0)$ and 
                                                       $V_1(x,a_0)$ 
associated with the supersymmetric Hamiltonians $H_0$ and $H_1$ such that
$$
H_s = -\frac{d^2}{dx^2} + V_s(x,a_0), \quad s = 0, 1,
$$ 
where $a_0$ is a set of real parameters. The shape 
invariance condition is defined by
$$
V_1(x,a_0) = 
V_0(x,a_1) + R(a_0),
\eqno (1)
$$
where $a_1 = h(a_0)$ corresponds to a reparametrisation in $V_0$ and $R(a_0)$ is 
a constant. The shape invariance condition immediately yields the energies 
and wavefunctions of $H_0$.$^{3,5}$ One obtain the 
energies $E_{n,0}$ of $H_{0}$
$$
E_{n,0} = \sum_{l=0}^{n-1} R(a_l), \quad  n \ge 1,
$$
where
$$
a_l = h^{(l)} (a_0) = h \circ h \circ \cdots \circ h (a_0), \quad 
l \ {\rm times}.
$$
(We take $E_{0,0} = 0$.) The three kinds of shape invariance mentioned in the 
introduction correspond to 
$a_1 = a_0 + \alpha$ with $\alpha \in {\bf R}$, 
$a_1 = \beta a_0$ with $0 < \beta < 1$, and 
$a_1 = h(a_0)$ with $h^{(l)}(a_0) = a_0$ for translational,$^{7,8}$ 
scaling,$^{9,10}$ and cyclic$^{11,12}$ shape invariance, respectively.

Another motivation for this work is to show that the isospectral Hamiltonians 
obtained from $k$-SUSYQM are connected through shape invariance. In this 
respect, we shall use some specific realizations of the $Z_k$-graded 
Weyl-Heisenberg 
algebra $W_k$ in order to generate a hierarchy of Hamiltonians 
subjected to translational or cyclic shape invariance.


\section{FRACTIONAL SUPERSYMMETRIC QUANTUM MECHANICS}



\subsection{Definition}


Let us go now to $k$-SUSYQM. A $k$-fractional supersymmetric dynamical system 
is defined by a triplet $(H, Q_+, Q_-)_k$ of 
operators satisfying the following relations$^{13-16,22}$ 
$$
H = H^{\dagger}, \quad 
Q_- = Q_+^{\dagger}, \quad 
Q_{\pm}^k = 0, 
$$
$$
\sum_{s=0}^{k-1} Q_-^{k-1-s}Q_+Q_-^s = Q_-^{k-2}H, \quad 
[H , Q_{\pm}] = 0,
\eqno (2)
$$
where $k = 3, 4, \cdots$. The Hamiltonian $H$ and the 
supercharges $Q_{\pm}$ of the dynamical system 
are linear operators acting on a Hilbert space 
${\cal H}$,
$$
{\cal H} = \bigoplus_{s=0}^{k-1} {\cal H}_s,
$$
which is $Z_k$-graded in view of the relations $Q_{\pm}^k = 0$. It 
is to be observed that Eq.~(2) works equally well in the case 
$k = 2$ corresponding to ordinary SUSYQM. 


\subsection{Generalized Weyl-Heisenberg algebra}


Following Ref.~22, we consider the generalized Weyl-Heisenberg 
algebra $W_k$, with $k \in {\bf N}\setminus\{0,1\}$, spanned by the four 
linear operators $X_+$, $X_-$, $N$ and $K$ acting on the space ${\cal H}$ and 
satisfying 
$$
X_- = X_+^{\dagger}, \quad 
N   =   N^{\dagger}, \quad
KK^{\dagger} = K^{\dagger}K = 1, \quad 
K^k = 1, \quad
$$
$$
[X_- , X_+] = \sum_{s=0}^{k-1} f_s(N)\Pi_s, \quad 
[N , X_-] = - X_-, \quad  
[N , X_+] =   X_+, 
$$
$$
KX_+ - q X_+K = 0, \quad 
KX_- - p X_-K = 0, \quad 
[K , N] = 0,
\eqno (3)
$$
where $q$ and $p$ are roots of unity with
$$
q = {\rm e}^{ \frac{2\pi {\rm i}}{k}}, \quad 
p = {\rm e}^{-\frac{2\pi {\rm i}}{k}}.
$$
In Eq.~(3), the functions 
$f_s: N \mapsto f_s(N)$ of the number operator $N$ are such that 
$$
f_s(N) = f_s(N)^{\dagger}.
$$
Furthermore, the operators 
$\Pi_s$ are defined in terms of the Klein or grading operator $K$ as 
$$
\Pi_s = \frac{1}{k} \sum_{t=0}^{k-1} p^{st} K^t, \quad s \in S_0.
$$
It is easy to check that 
$$
\Pi_s = \Pi_s^{\dagger}, \quad
\sum_{s=0}^{k-1} \Pi_s = 1, \quad 
\Pi_s \Pi_t = \delta_{s,t} \Pi_s.
$$
Consequently, the operators $\Pi_s$ are projection 
operators for the cyclic group $Z_k$. It can be proved 
that they satisfy 
$$
\Pi_s X_+ = X_+ \Pi_{s-1} \iff 
X_- \Pi_s = \Pi_{s-1} X_-.
$$
Note that the operators $X_+$ and $X_-$ can be considered as generalized 
creation and annihilation operators, respectively. 

It should be realized that, for fixed $k$, Eq.~(3) defines indeed a family 
of generalized Weyl-Heisenberg algebras $W_k$. The various members of the 
family are distinguihed by the various sets 
$\{f_s\} \equiv \{f_s(N) : s \in S_0\}$. 


\subsection{Realization of $k$-SUSYQM}


We can use the generators of $W_k$ for obtaining a realization of 
$(H, Q_+, Q_-)_k$.

First, we take the supercharge operators $Q_{\pm}$ (see Eq.~(2)) in the form 
$$ 
Q_- = X_- (1 - \Pi_1), \quad  
Q_+ = X_+ (1 - \Pi_0). 
\eqno (4)
$$
It can be proved that they satisfy the Hermitean conjugation property 
$Q_- = Q_+^{\dagger}$ and the $k$-nilpotency property 
$Q_{\pm}^k = 0$. Note that there are $k$ 
equivalent definitions of type (4) corresponding to the $k$ circular 
permutations of $0, 1, \cdots, k-1$.

Second, the $k$-fractional supersymmetric 
Hamiltonian $H$, satisfying (2) and compatible with (4), 
takes the form$^{22}$ 
$$ 
H = (k-1)X_+X_- 
 - \sum_{s=3}^{k}  \sum_{t=2}^{s-1} (t-1) f_t(N-s+t) \Pi_s
 - \sum_{s=1}^{k-1}\sum_{t=s}^{k-1} (t-k) f_t(N-s+t) \Pi_s, 
$$
in terms of $X_+ X_-$, $\Pi_s$ and $f_s$. In addition, 
it can be shown that the operator $H$ can be 
decomposed as 
$$
H = \sum_{s=1}^k H_s \Pi_s = 
    \sum_{s=0}^{k-1} H_{k-s} \Pi_{k-s}, 
\eqno (5)
$$
where
$$
H_s = (k-1)X_+X_- - \sum_{t=2}^{k-1} (t-1)f_t(N-s+t) + 
(k-1) \sum_{t=s}^{k-1} f_t(N-s+t), s \in S_1. 
\eqno (6)
$$
As an important result, it can be proved, from 
$[H , Q_{\pm}] = 0$, that the $k$ operators 
$H_k \equiv H_0$, $H_{k-1}$, $\cdots$, $H_1$ constitute a hierarchy
of isospectral Hamiltonians. Therefore, the spectra of 
$H_1, H_2, \cdots, H_{k-1}$ can be deduced from the spectrum of $H_0$. 


\subsection{Representation of $W_k$}


Let us now examine the action of $X_+$, $X_-$, $N$ and $K$ on each subspace
$$
{\cal H}_s = \{\vert n,s \rangle : n \ {\rm ranging}\}
$$
of ${\cal H}$ ($n$ can take a finite or infinite number of values according to 
whether as ${\cal H}_s$ is of finite or infinite dimension). For this purpose, 
we introduce the structure functions $F_s: N \mapsto F_s(N)$
through 
$$
X_+X_- = \sum_{s=0}^{k-1} F_s(N) \Pi_s, \quad 
X_-X_+ = \sum_{s=0}^{k-1} F_{s+1}(N+1) \Pi_s.
$$
In view of Eq.~(3), we have the recurrence relation 
$$
F_{s+1}(n+1) - F_s(n) = f_s(n), \quad  F_s(0) = 0. 
\eqno (7)
$$
Then, we can take$^{22}$ 
$$
X_+\vert n,s\rangle = \sqrt{F_{s+1}(n+1)}\vert n+1 ,s+1 \rangle, \quad  
s \ne k-1,
$$
$$
X_+\vert n,s\rangle = \sqrt{F_{s+1}(n+1)} \vert n+1,0 \rangle, \quad  
s = k-1,
$$
$$
X_-\vert n,s\rangle = \sqrt{F_s(n)}\vert n-1 ,s-1 \rangle, \quad  
s \ne 0, 
$$
$$
X_-\vert n,s\rangle = \sqrt{F_s(n)}\vert n-1,k-1\rangle, \quad  
s = 0,
$$
$$
N\vert n,s\rangle = n   \vert n,s\rangle, \quad
K\vert n,s\rangle = q^s \vert n,s\rangle 
\eqno (8)
$$
for the action of $X_+$, $X_-$, $N$ and $K$ on space ${\cal H}_s$. 
Relations (8) define a representation of $W_k$. 

In the following, we shall consider two special cases of $W_k$: 
(i) The case where $f_s(N)$ is independent of 
$s$ (see Section 4) and 
(ii) The case where $f_s(N)$ is independent of $N$ (see Section 5).


\section{TRANSLATIONAL SHAPE INVARIANT POTENTIALS}



\subsection{Structure function}


In this section, we assume that $f_s(N)$ is independent of $s$ and 
linear in $N$. More precisely, we take 
$$
f_s(N) = a N + b \Rightarrow [X_- , X_+] = a N + b, 
$$
with strictly positive eigenvalues, where $a$ and $b$ are two real 
parameters. Thus, from Eq.~(7) we have
$$
X_+X_- \equiv F(N,a,b),
$$ 
where
$$
F(N,a,b) = \frac{1}{2} a N(N-1) + b N.
$$
The non-linear spectrum of 
$X_+X_-$ is then given by
$$
X_+X_-\vert n,s\rangle = 
\bigg[\frac{1}{2} a n(n-1) + b n\bigg]\vert n,s\rangle.
$$

For either $a = 0$ and $b > 0$ or $a > 0$ and $b\geq 0$, the 
spectrum of 
$X_+X_-$ is infinite-dimensional
and does not present degeneracies. For $a < 0$ and $b \geq 0$, the spectrum 
of $X_+X_-$ is finite-dimensional with 
$n = 0, 1, \cdots, E(-\frac{b}{a})$ and all the states are non-degenerate.

It is possible to find a realization 
of each of the three cases just described 
in terms of an exactly solvable dynamical
system in a one-dimensional space, with coordinate $x$, and characterized 
by a potential $V(x, a, b)$. As a matter of fact, we have:

(i) $a = 0$ and $b = 1$ correspond to the harmonic oscillator potential
$$
V_{ho}(x, 0, 1) = x^2,
\eqno (9)
$$
with an infinite non-degenerate spectrum ($n \in {\bf N}$). 

(ii) $a = 2$ and $b = u+v+1$, with $u > 1$ and $v > 1$, correspond to the 
P\" oschl-Teller potential
$$ 
V_{PT} \bigg( x, 2, \{ u + \frac{1}{2}, v + \frac{1}{2} \} \bigg) = 
\frac{1}{4} \bigg[  
\frac{u(u-1)}{\sin^2\frac{x}{2}} + 
\frac{v(v-1)}{\cos^2\frac{x}{2}}\bigg] 
- \frac{1}{4} (u+v)^2,
\eqno (10)
$$
with an infinite non-degenerate spectrum ($n \in {\bf N}$). 

(iii) $a = -2$ and $b = 2l+1$, with $l \in {\bf N}$, correspond to the Morse 
potential
$$
V_{M}(x, -2, 2l+1) = {\rm e}^{-2x} - 
           (2l+3){\rm e}^{-x} + (l+1)^2,
\eqno (11)
$$
with an finite non-degenerate spectrum ($n = 0, 1, \cdots, l$). 


\subsection{Isospectral Hamiltonians}

 
The various isospectral Hamiltonians occuring in (5) are easily deduced 
from Eq.~(6). This gives 
 $$
 H_{k-s} \equiv H_{k-s}(N,a,b) = (k-1) \times
 $$ 
 $$
 \times \bigg[ F \bigg( N, a, b - \frac{1}{2} k a + a + s a \bigg) 
 + \frac{1}{6} (k-2) ( ka - 3 b )
 + \frac{1}{2}s(s-k+1)a + sb \bigg],
 \quad s \in S_0.
 $$
 Thus, the isospectral Hamiltonians are linked by 
 $$
 H_{k-s}(N,a,b) = H_0(N,a,b+sa) + \frac{1}{2} (k-1) s ( sa - a + 2b ), \quad
 s \in S_0,
 \eqno (12)
 $$
 a relation of central importance, in the $k$-SUSYQM context, for the 
 derivation of the translational shape invariance condition.

Let us denote by $V_k \equiv V_0$, $V_{k-1}$, $\cdots$, $V_1$ the potentials 
(in $x$-representation) associated 
with the isopectral Hamiltonians $H_0$, $H_{k-1}$,
$\cdots$, $H_1$, respectively. In other words, we set
$$
H_{k-s}(N,a,b) \equiv -\frac{d^2}{dx^2} + V_{k-s}(x,a,b), \quad 
s \in S_0.
$$
By using Eq.~(12), we immediately get the recurrence relation 
$$
V_{k-s}(x,a,b) = V_0(x,a,b+sa) + \frac{1}{2} (k-1) s ( sa - a + 2b ), \quad 
s \in S_0,
\eqno (13)
$$
which may be considered as the $k$-SUSYQM version 
of the translational shape invariance condition  
for ordinary SUSY (see Eq.~(1)).

By way of illustration, Eq.~(13) yields the following results.

(i) For the harmonic oscillator system:
$$
V_{k-s}(x,0,1) = x^2 + \frac{1}{2} (k-1)(2s - k + 2).
\eqno (14)
$$

(ii) For the P\" oschl-Teller system:
$$
V_{k-s} \bigg( x,2,\{ u + \frac{1}{2}, v + \frac{1}{2} \} \bigg) = \frac{1}{4}\bigg[
\frac{(u+s+1-\frac{k}{2})(u+s-\frac{k}{2})}{\sin^2\frac{x}{2}} +
\frac{(v+s+1-\frac{k}{2})(v+s-\frac{k}{2})}{\cos^2\frac{x}{2}}\bigg]
$$
$$
-\frac{1}{4}(u+v+2s+2-k)^2 +\frac{1}{6} (k-1)(k-2)(2k-3u-3v-3) + 
(k-1)s(s-k+u+v+2).
\eqno (15)
$$

(iii) For the Morse system:
$$
V_{k-s}(x,-2,2l+1) =  {\rm e}^{-2x} - (2l+k+1-2s){\rm e}^{-x}
+\frac{1}{4}(2l+k-2s)^2$$ $$-\frac{1}{6} (k-1)(k-2)(2k+6l+3)
+(k-1)s(k-s+2l)
\eqno (16)
$$
In the case  $k = 2$ and $s = 0$, Eqs.~(14), (15) and (16) 
                       reduce to Eqs.~(9),  (10) and (11), respectively.


\section{CYCLIC SHAPE INVARIANT POTENTIALS}



\subsection{Structure function }


In this section, we assume that $f_s(N)$ is independant of $N$, i.e.,
$$
f_s(N) = f_s \Rightarrow [X_- , X_+] = \sum_{s=0}^{k-1} f_s \Pi_s.
$$
(The paradigmatic case of the harmonic oscillator corresponds to 
$f_s = {\rm constant}$ for any $s$ in $S_0$.)

It is convenient to write the integer $n$ occurring in 
$\vert n,s\rangle$ as $n=kp+t$ with 
$p \in {\bf N}$ and $t \in S_0$. Here, to adapt our construction to
one-dimensional periodic potentials, we restrict the Hilbert-Fock 
space ${\cal H}$ to its subspace 
${\cal G} = \{ \vert kp+s,s \rangle : p \ {\rm ranging;} \ s \in S_0 \}$. In 
addition, it is appropriate to denote the state
$\vert kp+s,s \rangle$ as $\vert kp+s )$.
Hence, the action of the number operator $N$ on the 
states $\vert kn+s)$ is given by
$$
N \vert kn+s ) =(kn+s) \vert kn+s )
$$
and the grading operator $K$ can be identified, on 
the subspace ${\cal G}$, with the operator $q^N$ 
since
$$
K       \vert kn+s ) = 
q^s     \vert kn+s ) = 
q^{kn+s} \vert kn+s ) = 
q^N     \vert kn+s ).
$$

From Eq.~(7), it can be shown
$$
F_s(N) = g_0N + \sum_{t=1}^{k-1}g_t \frac{1-q^{st}}{1-q^t},
$$
where
$$
g_t = \frac{1}{k} \sum_{s=0}^{k-1} p^{st}f_s,  \quad  t \in S_0.
$$ 
Thus, the action of
$$
X_+X_- = 
\sum_{t=0}^{k-1} g_t \frac{1 - q^{Nt}}{1 - q^t}
$$
on the space ${\cal G}$ reads 
$$
X_+X_- \vert kn+s) = 
\bigg( n \sum_{i=0}^{k-1}f_i + \sum_{i=0}^{s-1}f_i \bigg)
       \vert kn+s). 
\eqno (17)
$$  

The spectrum of $X_+X_-$ is periodic and can be seen as a superposition of 
identical blocks. For a given block, the various gaps between the consecutive 
eigenvalues are 
$$
f_0, \ f_1, \ \cdots, \ f_{k-1}.
$$
The first block (corresponding to $n = 0$) has the following nonzero 
eigenvalues
$$
E_1 = f_0, \ 
E_2 = f_0 + f_1, \ \cdots, \ 
E_k = f_0 + f_1 + \cdots + f_{k-1}, 
$$
while the second block (corresponding to $n = 1$) has the eigenvalues 
$$
E_{k+1} = E_k + f_0, \ 
E_{k+2} = E_k + f_0 + f_1, \ \cdots, \ 
E_{2k } = E_k + f_0 + f_1 + \cdots + f_{k-1},  
$$
and so on for the subsequent blocks corresponding to $n = 2, 3, \cdots$ 
(the eigenvalue for the ground state is 
$E_0 = 0$). In other words, in the $(n+1)^{\rm th}$ block 
the parameter $f_s$ is the difference 
between the eigenvalues for $\vert kn+s+1)$ and 
$\vert kn+s)$. According to Eq.~(17), the various eigenvalues 
are given by
$$
E_{kn + s} = nk g_0 + \sum_{i = 0}^{s-1} f_i, \quad 
n \in {\bf N}; \ s \in S_0.
$$ 
Thus, each block has the length $kg_0$ which can be considered as 
the period of the cyclic spectrum.
 
At this level, it should be emphasized that our approach covers the one of 
Ref.~11
concerning the two-body Calogero-Sutherland model. The latter model 
corresponds to $k=2$. Consequently, the relevant Hilbert-Fock space is 
$$
{\cal G} = \{ \vert 2n+s ): n \in {\bf N}; s=0, 1\}
$$
and $X_+X_-$ reads 
$$
X_+X_- = \frac{1}{2}(f_0+f_1) N + \frac{1}{2}(f_0-f_1)\Pi_1.
$$
Equation (17) can then be particularized as 
$$
X_+X_- \vert 2n ) = n(f_0+f_1)\vert 2n ),
$$
$$
X_+X_- \vert 2n+1 ) = [n(f_0+f_1) + f_0]\vert 2n+1 ), 
$$
in accordance with the results of Ref.~11. (Our 
parameters $f_0$ and $f_1$ read $f_0 = \omega_0$ 
and $f_1=\omega_1$ in the notations of Ref.~11.)

It is interesting to note that that the spectrum of $X_+X_-$ coincides with 
one of the Hamiltonian corresponding to the potential (in $x$-representation) 
$$
V_0(x,f_0,f_1) = \frac{1}{16}(f_0+f_1)^2 x^2 + 
                 \frac{1}{4} \frac{(f_0-f_1)(3f_0+f_1)}{(f_0+f_1)^2} 
		 \frac{1}{x^2} -  
		 \frac{1}{2} f_1.
\eqno (18)
$$ 
Furthermore, using the standard tools of ordinary SUSYQM, we get 
$$
V_1(x,f_0,f_1) = \frac{1}{16}(f_0+f_1)^2 x^2 + 
                 \frac{1}{4} \frac{(f_1-f_0)(3f_1+f_0)}{(f_0+f_1)^2}
		 \frac{1}{x^2} + 
		 \frac{1}{2} f_0,
\eqno (19)
$$ 
which corresponds to the operator $X_-X_+$. 

For $k > 2$, the derivation of analytical forms of the potentials exhibiting 
a cyclic spectrum was discussed in Ref.~12.


\subsection{Isospectral Hamiltonians}


Going back to the general case, the expressions for the isospectral 
Hamiltonians in (6) can be obtained
from (5). This yields the relations
$$
H_{s} \equiv H_{s}(N, \{f_s\}) = (k-1) X_+X_- + \sum_{t=2}^{k-1} (1-t)f_t
+ (k-1) \sum_{t=s}^{k-1} f_t, 
\quad s \in S_1.
$$
These relations 
show that the spectra of the supersymmetric partner Hamiltonians 
$H_0$, $H_1$, $\cdots$, $H_{k-1}$ can be 
deduced from the one of $X_+X_-$ given by Eq.~(17). By combining 
the latter two relations, we obtain
$$
H_{k-s}(N,\{f_s\}) = H_0(N+s, \{f_s\}),  \quad  
s \in S_0, 
\eqno (20),
$$	  
an important relation  for the derivation of the cyclic shape invariance 
condition.

From (20), we can prove that
$$
H_{k-s}(N,\{f_s\}) = H_0(N, h^{(s)}\{f_s\}) + \sum_{i=0}^{s-1} f_i, \quad 
s \in S_0
\eqno (21)
$$
with
$$
h^{(s)} = h \circ h \circ \cdots \circ h, \quad s \ {\rm times}, 
$$
where $h$ is the circular permutation
$$
h \{f_s\} = h\{f_0, f_1, \cdots, f_{k-2}, f_{k-1}\} = 
\{f_1, f_2, \cdots, f_{k-1}, f_0\}
$$
such that $h^{(k)}$ is the identity.

We continue with dynamical systems in one-dimensional space 
(coordinate $x$). Let us note $V_0(x,\{f_s\})$ 
the potential associated with $H_0(N, \{f_s\})$:
$$
H_0(N, \{f_s\}) \equiv -\frac{d^2}{dx^2} + V_0(x,\{f_s\}).
$$
From Eq.~(21), it is easy to check that the potential $V_{k-s}(x,\{f_s\})$ 
associated with the Hamiltonian
$H_{k-s}(N, \{f_s\})$ can be obtained via
$$
V_{k-s}(x,\{f_s\}) = V_0(x,h^{(s)}\{f_s\}) + \sum_{i=0}^{s-1} f_i, \quad 
s \in S_0,
\eqno (22)
$$
to be compared with the cyclic 
shape invariance condition 
for ordinary SUSY (see Eq.~(1) and Refs.~11 and 12). 

As an example, for $k=2$, Eq.~(22) leads to 
$$
V_{1}(x , \{f_0, f_1\}) = 
V_{0}(x , \{f_1, f_0\}) + f_0,
$$
a relation satisfied by Eqs.~(18) and (19) 
for the Calogero-Sutherland potential.

\section{CONCLUDING REMARKS}
It was shown in the present paper how to tackle $k$-fractional SUSYQM through a 
$Z_k$-graded Weyl-Heisenberg algebra, noted $W_k$ with $k = 3, 4, \cdots$ (the 
case $k=2$ corresponding to ordinary SUSYQM). From the generators of this 
algebra, it was possible to find several realizations of $k$-fractional 
supersymmetric dynamical systems. Each system was characterized by a 
$k$-fractional supersymmetric Hamiltonian which gave rise to a hierarchy of $k$ 
isospectral Hamiltonians $H_{k-s}$ with $s \in S_0$. Two special 
cases of algebra $W_k$ were examined. They both led to $k$-fractional 
isospectral Hamiltonians, the potentials of which are connected by a 
recurrence relation that reflects a (translational or cyclic) shape invariance 
condition.

As a conclusion, $k$-fractional SUSYQM developed in the framework of algebra 
$W_k$ turns out to be a useful tool to generate a hierarchy of $k$ isospectral 
Hamiltonians linked by a translational or cyclic invariance condition.

A brief comparison with the results given by ordinary SUSYQM is in order. For 
$k=2$, the hierarchy of Hamiltonians reduces to a pair of isospectal 
Hamiltonians. Therefore, in order to generate a hierarchy of $k$ isospectal 
Hamiltonians, it is necessary to apply ordinary SUSYQM repeatedly. This is no 
longer the case for $k$-fractional SUSYQM since the hierarchy of $k$ 
isospectral Hamiltonians is generated at once. The equivalence between the 
approaches via ordinary SUSYQM applied repeatedly and $k$-fractional SUSYQM 
is ensured by the fact that $k$-SUSYQM can be seen as a superpostion 
of $k-1$ copies of ordinary SUSYQM.$^{22}$

To close this paper, let us offer two remarks. First, 
it is worthwhile to mention that our approach to 
$k$-SUSYQM by means of algebra $W_k$ can be applied to other
potentials. For instance, by taking $X_+X_- \equiv F(N, a, b, c)$,
where the structure function $F$ is given by
$$
F(N, a, b, c) = \frac{1}{2}aN(N-1) + bN + c \frac{1}{(N+1)^2}, 
$$
it might be possible to describe potentials involving a Coulombic 
part. Along this vein, a $k$-SUSYQM study to 
the effective screened potential,$^{40}$
singular inverse-power potentials,$^{41}$ and 
non-central potentials$^{42}$ could be fruitful. Second, 
it would interesting to examine the hidden supersymmetries 
exhibited by the Aharanov-Bohm, Dirac delta, and P\" oschl-Teller 
potentials$^{34-36}$ in the light of our approach to $k$-fractional SUSYQM. 
For this purpose, the connection between ordinary SUSYQM, possibly in a
$q$-deformed approach,$^{43}$ and $k$-SUSYQM$^{22}$ should play a central 
role.

\section*{Acknowledgment}

One of the authors (M.D.) would like to thank the {\em Groupe de 
Physique Th\'eorique} of the {\em Institut de Physique Nucl\'eaire de Lyon} 
for the kind hospitality extended to him during July 2006.


\newpage

\begin{thebibliography}{99}

\bibitem{1} J. Wess and B. Zumino, Nucl. Phys. B {\bf 70}, 39 (1974).
\bibitem{2} E. Witten, Nucl. Phys. B {\bf 188}, 513 (1981).
\bibitem{3} G. Junker, {\it Supersymmetric Methods in Quantum and Statistical Physics} 
(Springer, Berlin, 1996).
\bibitem{4} B. K. Bagchi, {\it Supersymmetry in Quantum and Classical Mechanics} 
(Chapman and Hall, London, 2000).
\bibitem{5} F. Cooper, A. Khare and U. Sukhatme, {\it Supersymmetry in Quantum 
Mechanics} (World Scientific, Singapore, 2001); 
Phys. Rep. {\bf 251}, 268 (1995).
\bibitem{6} L. Gendenshtein, JETP Lett. {\bf 38}, 356 (1983).
\bibitem{7} F. Cooper, J. N. Ginocchio and  A. Khare, Phys. Rev. D {\bf 36}, 2458 
(1987).
\bibitem{8} C. Chuan, J. Phys. A {\bf 24}, L1165 (1991).
\bibitem{9} A. Khare and U. Sukhatme, J. Phys. A {\bf 26}, L901 (1993).
\bibitem{10} D. T. Barclay, R. Dutt, A. Gangopadhyaya, A. Khare, A. Pagnamenta and 
U. Sukhatme, Phys. Rev. A {\bf 48}, 2786 (1993).
\bibitem{11} A. Gangopadhyaya and U. P. Sukhatme, Phys. Lett. A {\bf 224}, 5 (1996). 
\bibitem{12} U. P. Sukhatme, C. Rasinariu and A. Khare, Phys. Lett. A {\bf 234}, 401 
(1997).
\bibitem{13} V. A. Rubakov and V. P. Spiridonov, Mod. Phys. Lett. A {\bf 3}, 1337 
(1988).
\bibitem{14} J. Beckers and N. Debergh, Mod. Phys. Lett. A {\bf 4}, 1209 (1989); 
Nucl. Phys. B {\bf 340}, 767 (1990); 
N. Debergh, J. Math. Phys {\bf 34}, 1270 (1993); 
J. Phys. A. {\bf 26}, 7219 (1993); 
{\bf 27}, L213 (1994).
\bibitem{15} A. Khare, J. Phys. A {\bf 25}, L749 (1992); J. Math. Phys. {\bf 34}, 1277 
(1993).
\bibitem{16} A. T. Filippov, A. P. Isaev and A. B. Kurdikov, Mod. Phys. Lett. A 
{\bf 7}, 2129 (1992); Int. J. Mod. Phys. A {\bf 8}, 4973 (1993).
\bibitem{17} S. Durand, Mod. Phys. Lett. A {\bf 7}, 2905 (1992); Phys. Lett. B 
{\bf 312}, 115 (1993); Mod. Phys. Lett. A {\bf 8}, 2323 (1993).
\bibitem{18} S. Klishevich and M. S. Plyushchay, Mod. Phys. Lett. A {\bf 14}, 2379 
(1999).
\bibitem{19} M. Daoud, Y. Hassouni and M. Kibler, in {\it Symmetries in Science X}, 
edited by B. Gruber and M. Ramek (Plenum, New York, 1998);
Yad. Fiz. {\bf 61}, 1935 (1998).
\bibitem{20} M. Daoud and M. Kibler, in {\it Symmetry and Structural Properties of 
Condensed Matter}, edited by T. Lulek, B. Lulek and A. Wal (World Scientific, 
Singapore, 2001); in {\it Proceedings of the Sixth 
International Wigner Symposium} (Bogazici Univ. Press, Istanbul, Turkey, 2002).  
\bibitem{21} H.-Y. Pan and Z. S. Zhao, Phys. Lett. A {\bf 312}, 1 (2003).
\bibitem{22} M. Daoud and M. Kibler, Phys. Lett. A {\bf 321}, 147 (2004).
\bibitem{23} C. Ahn, D. Bernard and A. LeClair, Nucl. Phys. B {\bf 346}, 409 (1990); 
A. LeClair and C. Vafa, Nucl. Phys. B {\bf 401}, 413 (1993).
\bibitem{24} R. Kerner, J. Math. Phys. {\bf 33}, 403 (1992).
\bibitem{25} J. L. Matheus-Valle and M. A. R.-Monteiro, Mod. Phys. Lett. A {\bf 7}, 
3023 (1992); Phys. Lett. B {\bf 300}, 66 (1993); L. P. Collato
and J. L. Matheus-Valle, J. Math. Phys. {\bf 37}, 6121 (1996).
\bibitem{26} E. H. Saidi, M. B. Sedra and J. Zerouaoui, Class. and Quant. Gravity 
{\bf 7}, 1567 (1995).
\bibitem{27} N. Mohammedi, Mod. Phys. Lett. A {\bf 10}, 1287 (1995).
\bibitem{28} J. A. de Azc\'arraga and A. J. Macfarlane, J. Math. Phys. {\bf 37}, 1115 
(1996).
\bibitem{29} A. Perez, M. Rausch de Traubenberg and P. Simon, Nucl. Phys. B {\bf 482}, 
325 (1996); N. Fleury and M. Rausch de Traubenberg, Mod. Phys. Lett. A {\bf 11}, 
899 (1996); M. Rausch de Traubenberg and M. J. Slupinski, Mod. Phys. Lett. A 
{\bf 12}, 3051 (1997); M. Rausch de Traubenberg and P. Simon, Nucl. Phys. B 
{\bf 517}, 485 (1998); M. Rausch de Traubenberg and M. J. Slupinski, 
J. Math. Phys. {\bf 41}, 4556 (2000).
\bibitem{30} A. Mostafazadeh, Int. J. Mod. Phys. A {\bf 11}, 1057, 2941, 2957 (1996); 
K. Aghababaei Samani and A. Mostafazadeh, Nucl. Phys. B {\bf 595}, 467 (2001).
\bibitem{31} H. Ahmedov and \" O. F. Dayi, J. Phys. A {\bf 32}, 6247 (1999).
\bibitem{32} M. S. Plyushchay, 
Ann. Phys. {\bf 245}, 339 (1996); 
Mod. Phys. Lett. A {\bf 11}, 2953 (1996); 
Nucl. Phys. B {\bf 491}, 619 (1997);
Mod. Phys. Lett. A {\bf 12}, 1153 (1997).
\bibitem{33} M. Plyushchay, Int. J. Mod. Phys. A {\bf 15}, 3679 (2000).
\bibitem{34} F. Correa, M. A. del Olmo and M. S. Plyushchay, 
Phys. Lett. B {\bf 628}, 157 (2005).
\bibitem{35} F. Correa and M. S. Plyushchay, hep-th/0605104.
\bibitem{36} F. Correa, L.-M. Nieto and M. S. Plyushchay, hep-th/0608096.
\bibitem{37} M. Daoud and M. Kibler, Phys. Part. Nuclei (Supp. 1) {\bf 33}, S43 
(2002); Int. J. Quant. Chem. {\bf 91}, 551 (2003); 
M. R. Kibler and M. Daoud, in {\it Fundamental World of Quantum Chemistry}, 
edited by E. J. Br\"andas and E. S. Kryachko (Kluwer, Dordrecht, 2004). 
\bibitem{38} A. Chenaghlou and H. Fakhri, Int. J. Mod. Phys. A {\bf 18}, 939 (2003).
\bibitem{39} M. Sto\~ si\' c and R. Picken, math-phys/0407019. 
\bibitem{40} Shang-Wu Qian, 
Bo-Wen Huang and Zhi-Yu Gu, New J. Phys. {\bf 4}, 13.1 (2002).
\bibitem{41} B. G\"on\"ul, O. \"Ozer, M. Ko\c{c}ak, D. Tutcu 
and Y. Can\c{c}elik, J. Phys. A {\bf 34}, 8271 (2001).
\bibitem{42} M. Ko\c{c}ak, I. Zorba and B. G\"on\"ul, Mod. Phys. Lett. A
{\bf 17}, 2127 (2002).
\bibitem{43} A. N. F. Aleixo, A. B. Balantekin and M. A. C\^andido Ribeiro, 
J. Phys. A {\bf 36}, 11631 (2003).
\end{thebibliography}
\end{document}